\documentclass[10pt]{article}
\usepackage{psfig}
\usepackage{conf_iap}

\begin{document}

\heading{Large-Scale Structure from Wide-Field Surveys
} 
\par\medskip\noindent
\author{Nick Kaiser}
\address{
Institute for Astronomy, 
2680 Woodlawn Drive, 
Honolulu, HI 96822 \\
{\tt http://www-nk.ifa.hawaii.edu/}$\sim${\tt kaiser}
}

\begin{abstract} 
In this introductory talk I give an elementary overview of
the development, current status, and near 
term outlook for three probes of large scale structure; galaxy clustering,
bulk flows and weak lensing observations.
\end{abstract}

\section{The Paradigm for Structure Formation}

This meeting is largely concerned with the major observational
expansion occurring in the field of wide-field
cosmological surveys.  These advances are taking place on
a number of fronts, and promise to provide clear answers to a 
number of outstanding cosmological questions.
The fertility of this field owes much to the fact there has long been a
strong paradigm for structure formation; that the structure we see
in the Universe traces its roots back to small random fluctuations in
the early Universe which have evolved to the present by relatively
well understood physics.  While this picture may be wrong, and
has certainly undergone considerable evolution over time,
it has provided an important and 
influential framework for the interpretation of
cosmological observations.

The roots of this paradigm go back to Harrison \cite{harrison70}
and Zel'dovich \cite{zeldovich72}
who first seriously discussed the implications of the assumption
that structure originated in the big-bang; pioneering advances were
made by Peebles and others \cite{py70} in 
quantitative calculation of the evolution
of fluctuations from the radiation dominated era to the present, and
this `hierarchical clustering scenario' was developed to a refined state
by Gott and Rees \cite{gr75}
who painted a picture for the fluctuation
spectrum very similar in fact to the predictions from currently popular models.
Their `phenomenological' approach, working backwards from the
observed state of structure, was followed by the development
of inflationary scenario \cite{gs84} in which the Universe 
passes through a phase of
accelerated expansion which leaves the Universe 
in a state close to critical density. This is terminated by
re-heating \cite{kls97} and followed by the radiation era
and finally matter domination.  
 
On this background cosmology one can compute the evolution
of linearized fluctuations. It is a remarkable achievement
that this analysis allows one to follow a fluctuation with
present scale equal to that of a galaxy or cluster or
supercluster from when it was not much larger than the
Planck scale.  In models like `chaotic inflation'
these fluctuations start off as essentially massless 
zero-point fluctuations of the `inflaton' field
which, when they reach the horizon scale, get frozen in as fossilized 
ripples in the spatial
curvature \cite{gp82,starobinsky82,hawking82,bst83}, 
destined to re-enter the horizon
at much later times as classical density perturbations
with the so-called `Harrison-Zeldovich' spectrum.  The evolution
from the radiation to matter dominated eras and through decoupling
is solvable by integrating the Boltzmann equation \cite{be84,be87}.
At later time the evolution of fluctuations proceed
through the linear, quasi-linear ($\Delta \rho / \rho \sim 1)$,
and into the non-linear regime.  If the universe is dominated by
collisionless dark matter then the non-linear 
evolution can be calculated by
a variety of techniques, from direct N-body to analytic approximations.
The culmination of this evolution is shown by the Virgo
consortium simulation \cite{virgowebsite}.

Much of our knowledge of the large-scale structure comes from
the distribution of galaxies.  Here the otherwise clean
theoretical predictions for the matter distribution become
murky due to the possibility that the galaxy distribution
may be `biased' in some way.  
There are various types of bias that can arise.
One is the statistical `high peaks' bias.  This mechanism
explains the anomalously strong clustering
of clusters of galaxies \cite{k84b}, and has been explored
by analytic \cite{bbks86} and numerical \cite{co92}
methods.  The idea that galaxies may be biased in this
way is attractive, since it might provide a way to reconcile
dynamical estimates of $\Omega$ with a closed Universe.
The applicability of these results to real
galaxies however remains unclear, and there is the possibility of
competing effects from astrophysical biases such as ram
pressure stripping of gas from disk galaxies as they fall into clusters which would
tend to counteract the statistical bias effect.
In the face of these uncertainties, the common approach is
simply to parameterize the effect by a bias factor $b$, which
gives the ratio of density contrast of galaxies to that of the matter,
and treat this as a phenomenological parameter to be
fixed by observations.

This dark-matter dominated hierarchical clustering
paradigm has lasted well and accounts, quantitatively at least
for much of structure we see in the distribution of galaxies
today and at smaller scales at higher redshift in absorption
systems.  
The simplest version of the theory ($\Omega_m = 1$, CDM)
is now disfavored both from large-scale structure observations and
on other grounds \cite{gkc+98,henry97}, 
but there are various avenues for 
modification that have been fruitfully explored; one can modify the
early-universe physics and there have been a plethora of inflationary
models proposed, and one can modify the matter content by
incorporating hot dark matter for instance. Defects offer another
possibility, and the consequences of non-trivial
self-interaction of the dark matter have barely been explored.
The upshot of these models are predictions for the power spectrum
of mass density fluctuation $P(k)$
that can be tested against observations in
a number of ways.

\section{Galaxy Clustering}

Galaxy clustering is the most mature of the probes of LSS and,
for redshift surveys at least, gives a very direct way to probe the power
spectrum.  One can, for instance, compute the variance of galaxy
fluctuations counts in cells $\sigma^2$ as a function of cell, or smoothing, 
scale, and this is then related to the power spectrum by
\begin{equation}
\sigma^2 = b^2\int d^3k \; W^2(k) P(k)
\end{equation}
where $W(k)$ is the transform of the smoothing volume.
This is only illustrative, and more sophisticated techniques 
are actually used to measure $P(k)$ and estimate the measurement errors.

Low-redshift, large solid angle, redshift surveys 
such as the CfA, IRAS, LCRS and SSRS surveys
\cite{gh89,qdot90b,hg91,slo+96,dwp+98} have established a number
of facts: 
The power-law form for the 2-point function
$\xi_{\rm gg}(r) \propto r^{-\gamma}$ with $\gamma \simeq 1.8$ \cite{dp83,rgh90,tok+97}
corresponding to $P(k) \propto k^n$ with $n \simeq -1.2$, and the 
scaling of the 3- and 
4-point functions $\zeta$, $\eta$. 
The `cosmic virial theorem' \cite{dp83} result that if galaxies trace the mass
then $\Omega_m \simeq 0.2-0.3$.
The strong relative bias as a function of morphological type
\cite{dressler80,bts87,lmep95}.
The large-scale filamentary or web-like inter-cluster structure, which
resonates with the theoretical picture \cite{bkp96}.  
The power spectrum has been measured \cite{fkp94,dvg+94,pvgh94,lks+96} 
and seems to continue
rising with wavelength to
$\lambda \sim 200 h^{-1}$Mpc, in conflict with standard CDM predictions,
beyond which the data give out.
There have been attempts to measure the redshift-space distortion effect 
\cite{hamilton93,lemp96}
but these have been somewhat noisy to date.

Deeper pencil beam surveys 
show a remarkable regularity of the distribution of clumps
of galaxies \cite{beks90}, and there have been suggestions
that the large scale structure in the LCRS \cite{lsl+96} also
appears non-Gaussian, though see \cite{kp92}. The CFRS survey \cite{cfrs96}
shows strong negative evolution of the clustering strength going back to
$z \simeq 1$, which has been compared to theoretical predictions
by Peacock \cite{peacock97}, though this 
to be contrasted with the strong clustering found by
Steidel and colleagues \cite{sad+98,asg+98}
in their $z \sim 3$ color selected survey 
where we are perhaps seeing statistical `high-peaks' biasing 
of galaxies in action.

Angular surveys provide an important complement to $z$-surveys; 
the clustering is smeared
out somewhat in projection, but the great volume of these surveys makes them
potentially valuable probes of $P(k)$ beyond the expected peak in $P(k)$.
The 2-D power spectrum is simply related to the 3-D spectrum by
a convolution in log frequency space \cite{k92b}, and deconvolution 
of the measured angular power spectrum \cite{be93}
suggests that we are already seeing the turnover.

In the near term future the big developments in relatively low redshift,
large solid angle surveys will be the 
Sloan SDSS \cite{gw95,sdsswebsite} and 
Anglo-Australian 2dF \cite{taylor95,2dfwebsite}
surveys.  These should probe the power spectrum to lower $k$ and should
convincingly reveal the turnover in $P(k)$ if it exists.  They should
give the signal to noise required to measure the $z$-space distortion effect.   
It should be possible to better quantify the
dependence of galaxy clustering on galaxy type, and the higher precision for
higher order statistics open up the possibility of testing for
dynamical and bias associated non-gaussianity \cite{cbbh97}, 
and may also provide
constraints on non-gaussianity of the primordial fluctuations \cite{sp96}. 
The successor to the LCRS survey will be the DEEP survey,
which will explore a slice of the Universe at $z \sim 1$, allowing
detailed study of the evolution of clustering, and $z$-space
distortion effect \cite{k87}, including the asphericity due to
the cosmological background \cite{ap79,ryden95}.
State of the art
high redshift surveys \cite{cshc96} extend to magnitudes $m_I \sim 23$, at which point
spectra take several hours on a 10m class telescope to obtain and there
are severe problems in getting firm redshifts for many faint galaxies
at $z \sim 2$.  
The VIRMOS survey, with its 120 VLT nights,  should 
however make a significant impact here. Clustering studies from
angular surveys will benefit from the development of
wide format CCD cameras \cite{lmkcgm95,lts98} 
and will further benefit from photometric redshift
estimation \cite{bcsb97,sly97,fly98}
which will help the deprojection.
Many of the talks in this volume expand on these points.  

\section{Bulk-Flows}

Bulk-flow studies developed somewhat later.  The idea here is that there should
be distortion of the cosmic expansion field 
associated with the growth of structure, so
if one can determine accurate redshift independent distances to galaxies then one can
measure this `peculiar velocity field' 
(for reviews see \cite{k90b,k90a,sw95,dekel94}).  
This provides a probe of $P(k)$ with the variance in 
velocity smoothed on some scale being
\begin{equation}
\sigma_v^2 = \int d^3k \; (H/k)^2 W^2(k) P(k)
\end{equation}
Advantages of this technique are that it
provides a direct probe of the mass fluctuations
independent of bias, and also 
(essentially due to the $1/k^2$ weighting in the integral here) 
provides a nice probe of
fluctuations in the linear and quasi-linear regimes.
The disadvantage is that the technique is limited to the 
fairly nearby Universe
because of the nature of the distance errors.  Bulk flow $P(k)$ measurements 
\cite{zzd+97,kd97} are therefore
subject to substantial sampling, or `cosmic', variance. 
These observations can however also be
profitably used to compare the inferred mass 
fluctuations with the
galaxy distribution (by POTENT or other techniques \cite{bd89,ks90}), and
in this way test the gravitational instability picture.

The early history of bulk-flow measurements was at times
confusing and contradictory,
but by the late 80's a fairly coherent picture emerged \cite{abc+89,dfb+87}
from samples of typically several hundreds of galaxies, of substantial
flows on fairly large-scales, and that the 600 km/s motion of the local
group with respect to the microwave background radiation is neither atypical 
nor a small-scale local fluctuation.  The divergence of these
flows seems to (anti) correlate with the galaxy distribution \cite{qdot91,dby+93}
as expected in the gravitational instability picture.  
Most of these comparative studies have concentrated on IRAS galaxies, and seem to
indicate a rather high value of $\Omega$ \cite{dekel94}, subject to the question of 
bias, though see \cite{dr94}.  
This result is supported
by the `dipoles' analysis \cite{kl89} which seems to suggest that
the acceleration of the local group agrees in direction with the observed motion, and,
if the acceleration has converged, that $\Omega$ is high.  One should
not discount here the possibility of additional attraction from
large scales \cite{svz94}.  The `7-samurai' study \cite{dfb+87} led famously to the
discovery of the Great Attractor \cite{lfb+88,bfd90,bdf+90}; which we take to be the
conclusion that there is a large mass concentration lying behind
Centaurus cluster -- this cluster's large apparent motion with respect to to the MBR not being
predicted from the distribution of IRAS galaxies at least \cite{qdot91}.

More recently, and with much effort, the sample sizes have been increased
to thousands of galaxies \cite{mf94,wcf+97}.   The comparison 
\cite{dnw96} of the MkIII velocities
with the IRAS galaxy distribution revealed some rather worrying systematic
discrepancies, including flow patterns of a form hard to reconcile with
gravitational instability, but the comparison of the IRAS galaxy distribution
with the SFI velocity dataset
\cite{dnf+98} seem to show
better agreement, and seem to further support a high density Universe
$\beta \equiv \Omega^{0.6} / b \simeq 0.6$ in line with other studies above. 
Brightest cluster galaxies \cite{lp94} provide an attractive way to
increase the sample depth, but the results remain somewhat controversial.
An important development taking place is the augmentation of the
conventional Tully-Fisher, Faber-Jackson, $D_n-\sigma$ techniques
with distances from surface-brightness fluctuations \cite{tbad97}
which give greatly reduced
statistical uncertainty, and 1a supernovae distances \cite{gkc+98}
may also provide
a useful check on the other techniques.

In contrast to the apparent strong flows on large-scales the 
small scale flow field is remarkably cold \cite{sandage86,bp87}.  This has
been characterized as the `cosmic Mach number problem' \cite{os90}
and remains a challenge to cold-collisionless dark matter models.

\section{Weak Lensing}

Weak lensing, in the sense of the statistical distortion of the shapes of 
faint background galaxies, has now been measured for
quite a number of clusters:
\cite{twv90,bfkms93,fksw94,bmf94,dml94,mdfb94,fm94,sd95,tf95,fmdbk96,skss96,bs97,sedcosb97,fbrt97,ft97},
and provides a direct measurement of the
total mass distribution in clusters.
Following the pioneering attempt of Tyson and colleagues \cite{tvjm84}
using photographic plates,  several groups 
\cite{bbs96,dt96,gcir96,hsdk98})
have reported CCD measurements of the `galaxy-galaxy lensing' effect
due to dark halos around galaxies,
and there have also been estimates of the shear due to large-scale structure
\cite{vjt83,mbvbssk94,svmjsf98}, the shear variance being related to
the power spectrum by
\begin{equation}
\sigma_\gamma^2 \sim (H_0 D/c)^3 \int d^3k \; (H/k) P(k)
\end{equation}
and there has been much
theoretical activity in prediction of large-scale shear
\cite{bsbv91,bvm97,gb98,js97,miraldaescude91,k92b,waerbeke98,villumsen96}
and in reconstruction techniques \cite{ks93,sk96,bnss96,lb98b,ss95b}.
Most of the observational studies have been made with fairly small
CCD detectors; this severely limits the distance out to which one
can probe the cluster mass distribution and also limits the precision
of galaxy-galaxy lensing and large-scale shear studies. 

Recently we have presented a weak lensing analysis 
\cite{kwl+98} of deep $I$ and $V$ photometry of the 
field containing the $z \simeq 0.42$ supercluster
MS0302+17 taken with a large 8192$\times$8192 pixel CCD mosaic camera,
the UH8K \cite{lmkcgm95}, mounted behind the prime focus
wide field corrector on the CFHT.  
The field of view of this camera on this telescope measures
$0^\circ.5$ on a side, and greatly increases the range of accessible scales.

\begin{figure}[htbp!]
\centerline{\vbox{\psfig{figure=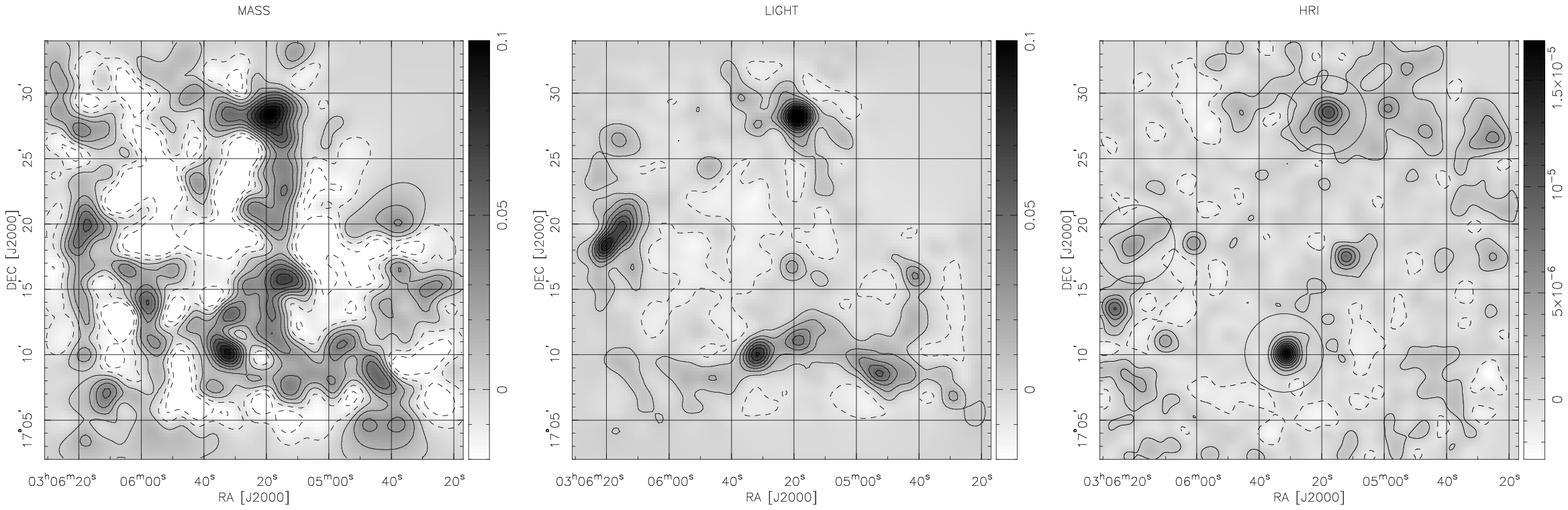,height=4cm}}}
\caption[]{Left hand panel shows the 2-D mass reconstruction.  The center
panel shows the predicted $\kappa$ due to structures at $z \sim 0.2-0.6$
assuming that early type galaxies trace the mass.  Right hand panel shows
a smoothed X-ray image from the ROSAT HRI; sources other than those circled appear
to be point-like in the unsmoothed image and are most likely unassociated with the
supercluster.
}
\label{fig:mlx}
\end{figure}

We assembled the $\sim 5$hrs of multiple exposure into a composite image
on which we detected $\sim 40,000$ faint galaxies whose shapes we
measured to obtain estimates of the weak lensing shear field $\gamma(r)$
and thereby the dimensionless surface density $\kappa(r)$, the
ratio of the physical surface density to the critical density, as shown in
the left panel of figure \ref{fig:mlx}.

\begin{figure}[htbp!]
\centerline{\vbox{\psfig{figure=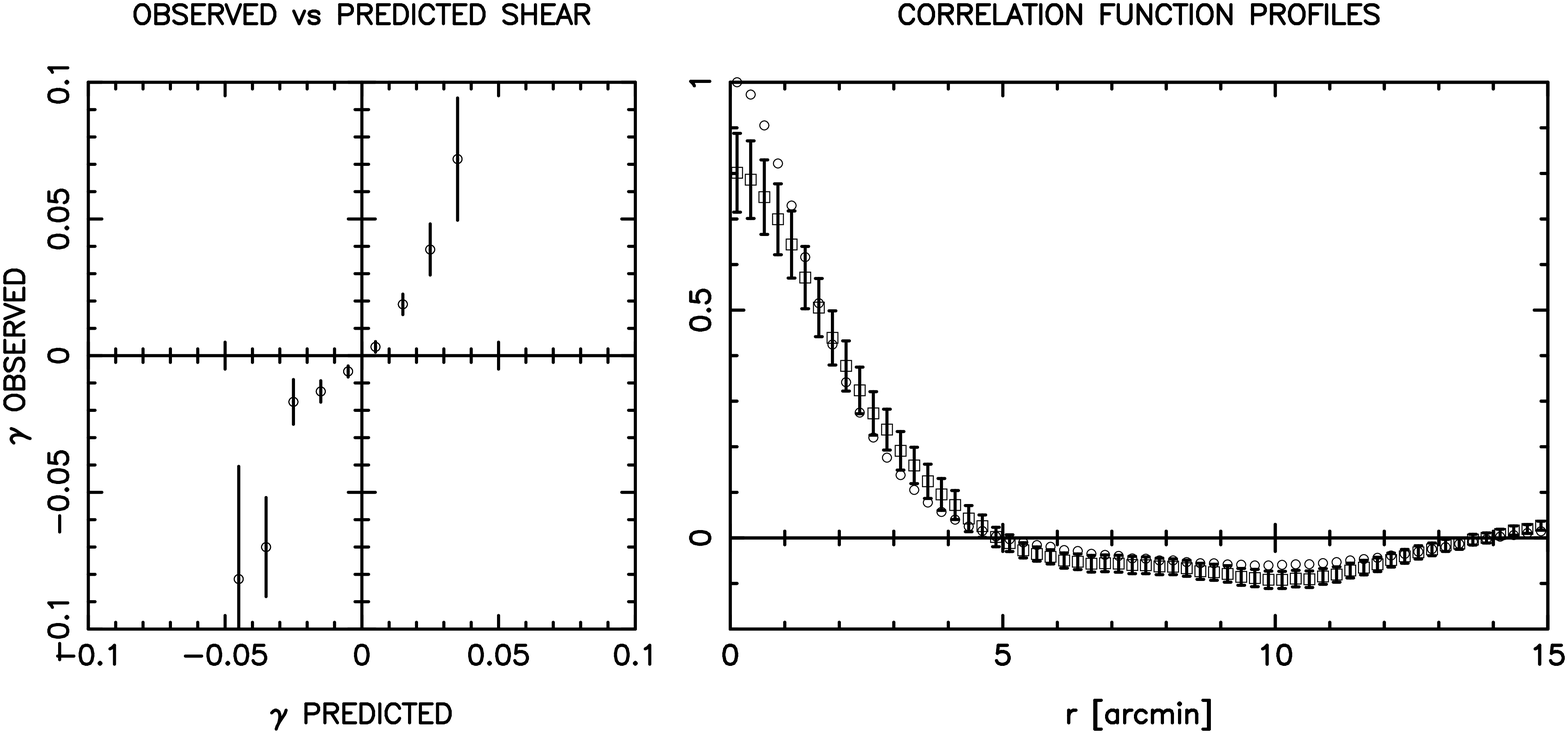,height=5cm}}}
\caption[]{Left panel shows the strong correlation between the
measured shear and that predicted, here computed assuming 
early type galaxies trace the mass
with a nominal mass-to-light ratio $M/L = 200h$ and assuming a mean effective
redshift for the background galaxies of $z = 1.5$. Fitting for a linear
relation gives
$M/L \simeq 250 h$.  Symbols
with error bars in the right hand panel shows the cross correlation
of the measured surface density with the luminosity density
(for $M/L = 280 h$) and the circles show the auto-correlation of the
luminosity surface density.
}
\label{fig:shearcorrelation}
\end{figure}

Our hope was to compare the mass distribution with the distribution of
galaxy light in the supercluster, but measuring the latter proved to be difficult due
to the presence of foreground clusters.  The elliptical galaxies in the 
supercluster and other high-$z$ structures are however readily separable from the
foreground clutter and we were able to generate a prediction for $\kappa(r)$
assuming that the mass is distributed like these E-galaxies.  We did not expect 
this to agree with the measured mass; E-galaxies are known to be strongly
clustered and concentrated in the densest parts of clusters, with the bulk of
the population having more extended profiles, and in high density biased
models the dark matter would be yet more extended.  To our surprise the
predicted and observed shear agreed remarkably well (see figure \ref{fig:mlx}).
There is considerable noise in the reconstructions, but we can make a fairly
precise estimate of the cross-correlation between the mass and the light
as shown in figure \ref{fig:shearcorrelation}.  The mass-to-light ratio
we measure on the largest scales ($\lambda \sim 6 h^{-1} {\rm Mpc}$) is 
$M/L_B \simeq 280 h$, and is very similar to that we find for the centers of the
clusters, and is considerably lower than has been found for
ultra-massive clusters like A1689 \cite{tf95}.  The clusters in MS0302+17 do
not have extended halos, and the mass follows the E-galaxy profile
very accurately.   

Exactly what these results imply for the cosmological density parameter
remains somewhat unclear due to to the possibility of bias, but, at
face value, indicate that there is very little mass associated 
with late type galaxies, that the density parameter is very
low $\Omega \simeq 0.05$, and that much or perhaps all of the dark matter is
baryonic.

\acknowledgements{I wish to thank the Stephane and Yannick 
for organizing a great meeting.}

%
%
%



\def\aj{{AJ}}                   
\def\araa{{ARA\&A}}             
\def\apj{{ApJ}}                 
\def\apjl{{ApJ}}                
\def\apjs{{ApJS}}               
\def\ao{{Appl.~Opt.}}           
\def\apss{{Ap\&SS}}             
\def\aap{{A\&A}}                
\def\aapr{{A\&A~Rev.}}          
\def\aaps{{A\&AS}}              
\def\azh{{AZh}}                 
\def\baas{{BAAS}}               
\def\jrasc{{JRASC}}             
\def\memras{{MmRAS}}            
\def\mnras{{MNRAS}}             
\def\pra{{Phys.~Rev.~A}}        
\def\prb{{Phys.~Rev.~B}}        
\def\prc{{Phys.~Rev.~C}}        
\def\prd{{Phys.~Rev.~D}}        
\def\pre{{Phys.~Rev.~E}}        
\def\prl{{Phys.~Rev.~Lett.}}    
\def\pasp{{PASP}}               
\def\pasj{{PASJ}}               
\def\qjras{{QJRAS}}             
\def\skytel{{S\&T}}             
\def\solphys{{Sol.~Phys.}}      
\def\sovast{{Soviet~Ast.}}      
\def\ssr{{Space~Sci.~Rev.}}     
\def\zap{{ZAp}}                 
\def\nat{{Nature}}              
\def\iaucirc{{IAU~Circ.}}       
\def\aplett{{Astrophys.~Lett.}} 
\def\apspr{{Astrophys.~Space~Phys.~Res.}}
\def\bain{{Bull.~Astron.~Inst.~Netherlands}} 
\def\fcp{{Fund.~Cosmic~Phys.}}  
\def\gca{{Geochim.~Cosmochim.~Acta}}   
\def\grl{{Geophys.~Res.~Lett.}} 
\def\jcp{{J.~Chem.~Phys.}}      
\def\jgr{{J.~Geophys.~Res.}}    
\def\jqsrt{{J.~Quant.~Spec.~Radiat.~Transf.}}
\def\memsai{{Mem.~Soc.~Astron.~Italiana}}
\def\nphysa{{Nucl.~Phys.~A}}   
\def\physrep{{Phys.~Rep.}}   
\def\physscr{{Phys.~Scr}}   
\def\planss{{Planet.~Space~Sci.}}   
\def\procspie{{Proc.~SPIE}}   

\let\astap=\aap
\let\apjlett=\apjl
\let\apjsupp=\apjs
\let\applopt=\ao

\end{document}